\documentclass[authorversion=true, anonymous=false, nonacm=true]{acmart}

\renewcommand\footnotetextcopyrightpermission[1]{} % removes footnote with conference information in first column

\AtBeginDocument{%
  \providecommand\BibTeX{{%
    \normalfont B\kern-0.5em{\scshape i\kern-0.25em b}\kern-0.8em\TeX}}}

\setcopyright{none}
\acmJournal{PACMHCI}
\acmYear{2021} \acmVolume{x} \acmNumber{x} \acmArticle{x} \acmMonth{} \acmPrice{15.00}\acmDOI{10.1145/3359276}

\usepackage{cleveref}
\usepackage{lscape}
\usepackage{graphicx}
\graphicspath{ {./images/} }
\usepackage{enumitem}
\usepackage{booktabs}
\usepackage{multirow}

\newcommand{\newwriting}{\textcolor{black}}

\newcommand{\newer}{\textcolor{black}}

\newcommand{\newest}{\textcolor{black}}

\begin{document}

% \title[Clubhouse and MIC]{Understanding moderation on Clubhouse using MIC: \\ A New Affordance-aware Framework for Navigating Moderation in the Cacophony of Audio-based Social Platforms}
\title[MIC: Affordance-Aware Framework for Platform Moderation]{Harmonizing the Cacophony with MIC: An Affordance-aware Framework for Platform Moderation}

\author{Tanvi Bajpai}
\email{tbajpai2@illinois.edu}
\affiliation{%
  \institution{University of Illinois at Urbana-Champaign}
%  \streetaddress{P.O. Box 1212}
  \city{Urbana}
  \state{Illinois}
  \country{USA}
  \postcode{61801}
}

\author{Drshika Asher}
\email{drshika2@illinois.edu}
\affiliation{%
  \institution{University of Illinois at Urbana-Champaign}
%  \streetaddress{P.O. Box 1212}
  \city{Urbana}
  \state{Illinois}
  \country{USA}
  \postcode{61801}
}

\author{Anwesa Goswami}
\email{anwesag2@illinois.edu}
\affiliation{%
  \institution{University of Illinois at Urbana-Champaign}
%  \streetaddress{P.O. Box 1212}
  \city{Urbana}
  \state{Illinois}
  \country{USA}
  \postcode{61801}
}

\author{Eshwar Chandrasekharan}
\email{eshwar@illinois.edu}
\affiliation{%
  \institution{University of Illinois at Urbana-Champaign}
%  \streetaddress{P.O. Box 1212}
  \city{Urbana}
  \state{Illinois}
  \country{USA}
  \postcode{61801}
}

%%%for review
\makeatletter
\let\@authorsaddresses\@empty
\makeatother


\renewcommand{\shortauthors}{Bajpai et al.}

% \newcommand{\remark}[]{\textcolor{red}}
% \newcommand{\temp}[]{\textcolor{blue}}
\begin{abstract}
\newest{Social platforms, and the online communities that use them, are evolving at a rapid pace. As a result, research and development regarding how to moderate online communities is being out-paced.} In this paper, we present a novel framework that \newest{will allow moderation researchers and practitioners to not only keep-up with the diverse landscape of available platforms and affordances, but also comprehensively represent and analyze moderation on these platforms.} \newest{The MIC framework represents a social platform's \emph{moderation ecosystem}} using a base-set of 12 platform-level affordances, \newest{along with a notion of the} inter-affordance relationships that can exist between them. These affordances fall into the three categories---\textbf{M}embers, \textbf{I}nfrastructure, and \textbf{C}ontent\newest{---that are derived from Grimmelmann's taxonomy of moderation.}
% \cite{grimm}, a framework that is already widely accepted and used by the moderation research community.} 
% \newest{To show how MIC serves as an insightful augmentation of Grimmelmann's lens, we begin by describing how its components have already been shown to impact Grimmelmann's techniques for moderation. Then, }
We demonstrate the advantages of using an affordance-aware framework like MIC by analyzing several social platforms \newest{over the course of} two case studies. First, we analyze individual platforms using MIC and demonstrate how MIC can be used to examine the effects of platform changes on the moderation ecosystem and identify potential new challenges in moderation. Next, use MIC to systematically compare three platforms and propose potential moderation mechanisms that each can adapt. Moderation researchers and \newest{stakeholders} can use such comparisons to uncover where platforms can emulate established, successful and better-studied platforms, as well as learn from the pitfalls other platforms have encountered.
% We conclude by discussing implications and advantages of using the MIC framework.
%   Clubhouse is an audio-based social platform that launched in April 2020 and rose to popularity amidst the global COVID-19 pandemic. Unlike other platforms such as Discord,  Clubhouse is entirely audio-based, and is not organized by specific communities. Following Clubhouse's surge in popularity, there has been a rise in the development of other audio-based platforms, as well as the inclusion of audio-calling features to existing platforms.
%   In this paper, we present a framework (MIC) for analyzing audio-based social platforms that accounts for unique platform affordances, the challenges they provide to both users and moderators, and how these affordances relate to one another using MIC diagrams. 
%   Next, we demonstrate how to apply the framework to preexisting audio-based platforms and Clubhouse, highlighting key similarities and differences in affordances across these platforms.
%  Using MIC as a lens to examine observational data from Clubhouse members we uncover user perceptions and challenges in moderating audio on the platform.
  
\end{abstract}





\maketitle

\section{Introduction}

The moderation of online communities has been the focus of a large body of social computing research \cite{kiesler2012regulating,jhaver18, myers2018censored, gilbert2020run,seering17,seering2019moderator,roberts2016commercial,lampe04,jiang19,kiene16,kiene19,schlesinger17, chandrasekharan2017you}. Much of this research is unified by the use of Grimmelmann's taxonomy of moderation \cite{grimm}, which defines, \newest{in generality, the terminology, goals, techniques, and challenges of moderating online communities.} Grimmelmann characterizes an online community \newest{using three features}: the community's \emph{members}, the \emph{content} that is shared among the members, and the \emph{infrastructure} used to share it. Grimmelmann's four techniques for moderation, \emph{excluding, pricing, organizing and norm-setting,} are all defined in a way that is \newest{agnostic of the diverse communities and technologies that will implement them. For example, \emph{exclusion} refers to keeping unwanted members from joining the community. On Reddit, this can be done by banning problematic users from a community (subreddit) page~\cite{jhaver2019human}. On Discord, community (server) administrators may require members to verify that they have a domain-specific email address; to do so, they can create custom authentication applications using Discord developer tools.\footnote{An example of one such Discord tool can be found at \url{https://github.com/sigpwny/sigpwny-shibboleth-auth}}}

% The \newest{platform-agnostic nature} of 
Grimmelmann's taxonomy is unequivocally useful for unifying moderation \newest{across communities and technologies. However, in its current formulation, Grimmelmann's taxonomy does not explicitly account for nuances at the platform-level (e.g., affordances, technological or design related).
It is clear that the Social Networking Sites (SNSs), or social \emph{platforms}, that online communities use shape many of the moderation challenges they face, as well as the strategies they employ to address them. In particular, it is the platform's \emph{moderation ecosystem}, i.e. the components that impact moderation and the interactions among them, that play a central role in how the communities that use them are moderated.} As such, \newest{recent} moderation research is centered around particular platforms (e.g.,~\cite{jiang19,schlesinger17}), or newer technologies such as Virtual Reality (VR)~\cite{blackwell2019harassment}. As more platforms are created and updated, so too are the moderation strategies, needs, and challenges of the online communities that use them. For instance, communities that use voice rooms \cite{jiang19} or VR technology \cite{blackwell2019harassment} face novel moderation challenges unlike those of communities on more traditional and better-studied platforms such as Reddit or Facebook. \newest{Hence, it is incredibly important to research and develop moderation on newer platforms as they emerge and become more popular. }

% However, the landscape of online communities and moderation work on these communities is being rapidly out-paced by the development of platforms.

% \newwriting{The landscape of Social Networking Sites, or social \emph{platforms}, has been growing rapidly since the birth of the internet. Furthermore, existing social platforms keep adding and updating features as new technology and trends develop. As such, it is more important than ever to understand how the online communities on these platforms are and should be moderated. While there is already a large body of work done on understanding moderation online \cite{kiesler2012regulating,jhaver18, myers2018censored, gilbert2020run,seering17,seering2019moderator,roberts2016commercial,lampe04,jiang19,kiene16,kiene19,schlesinger17, chandrasekharan2017you}, the acceleration of platform development is making it more challenging to keep moderation research up-to-date and ensure that state-of-the-art moderation solutions do not cause more moderation challenges on new or updated platforms.}

\newest{Unfortunately, platform development typically outpaces moderation research.} A clear example of this can be seen in the recent rise in popularity of audio-based social platforms \newest{such as Clubhouse, which launched into the mainstream during the global COVID-19 pandemic \cite{solans2020rise}.}
%In March of 2020, the global COVID-19 pandemic forced people to self-isolate, work from home, and limit in-person interactions all together; this allowed for a new social platform called Clubhouse to surge into the mainstream \cite{solans2020rise}. 
Clubhouse's success was closely followed by the introduction of other audio-focused platforms and extensions to existing platforms \cite{radcliffe2021audio}. \newest{Sonar, an alternative voice-chatting app, launched in January 2021 \cite{sonar_2021}. Both Twitter and Facebook launched live audio room features during 2021. Other popular platforms such as Reddit \cite{peters_2021}, Telegram \cite{wilson_2021}, Slack \cite{sarwar_2021}, and Discord \cite{tech2_2021} quickly followed suit and began launching their own Clubhouse-esque features. During this time, Spotify acquired the parent company of an audio-only, sports-centered app called Locker Room \cite{spotify_2021,steele_2021}, and later re-branded and re-launched it as direct competitor to Clubhouse called Spotify Greenroom \cite{carman_2021}. A timeline of this audio-based platform ``boom'' can be found later on in \Cref{fig:tl-absps}.}

% Platforms for audio conversations have existed prior to Clubhouse's creation. Many video games have supported in-game voice chat since the early 2000s \cite{loguidice2014vintage}, and applications such as WhatsApp\footnote{\url{https://www.whatsapp.com/}} and Discord\footnote{\url{https://discord.com/}} have supported audio communication since their conception, along with text-based communications. Clubhouse, on the other hand, is an \textit{audio-only} platform, and provides no way besides voice for its users to communicate inside the platform. 

% to uncomment
% \begin{figure}
%     \centering
%     \includegraphics[width =\textwidth]{images/timeline.pdf}
%     \caption{A timeline of popular audio-based technologies and social platforms. Clubhouse appears to mark the beginning of an audio-based ``boom'' in platform development.}
%     \label{fig:tl-absps}
% \end{figure}

\subsection{Moderation Research in the Landscape of Evolving Social Platforms}
Similar to the development of any new social technology, questions about moderating such platforms continues to be of particular interest to not only the Computer-Supported Cooperative Work and Social Computing (CSCW) research community, \newest{but also other moderation stakeholders such as platform designers, moderators, and community members}. We identify three key challenges that researchers \newest{and stakeholders face when addressing} moderation in the landscape of dynamically evolving social platforms.
First, it may be tempting to choose one or two representative platforms \newest{while developing} new insights to their moderation \newest{when,} in reality, these platforms are diverse in ways that effect moderation. For instance, \newest{despite their similarities,} Spotify Greenroom allows users to enable a text-based chat box in their live audio room while Clubhouse does not. Secondly, many of the new platforms or features might appear to be novel or unstudied, when they are \newest{essentially repackaged versions of older and more established technologies.} Spotify Greenroom's chat boxes are similar to those from Twitch, a popular live video-streaming platform; Sonar's world-building concept resembles classic virtual world building games such as Minecraft. \newest{And finally, as was previously discussed, these platforms are rapidly evolving and adding features that not only impact moderation but also could undermine or out-date moderation research by the time it gets published or implemented. For instance, Clubhouse added several new features that impact moderation during the time between this manuscript's submission and publication.}  

\newwriting{To address these challenges, and better enable the moderation research community to keep up with rapid platform development, we develop a new theoretical framework to \newest{aid in the analysis of moderation} on social platforms. Our framework can benefit \newest{moderation stakeholders} by enabling them to identify potential moderation challenges they could face when using platform, as well as \newest{adapt or} design moderation solutions to address them.}

\subsection{The MIC Framework}
% The contributions in this paper are three-fold.
% First, we present a novel platform-aware framework for understanding moderation on audio-based social platforms.

In this paper, we present a novel theoretical framework that allows us to represent social platforms' \emph{moderation ecosystems}. 
%By moderation ecosystem, we mean the physical attributes of a social platform that impact moderation.  
\newest{This representation is comprised of a} base set of twelve relevant platform-level \emph{affordances}. Each affordance falls into one of three categories that are derived from Grimmelmann's~\cite{grimm} definition of an online community: \textbf{M}embers, \textbf{I}nfrastructure, and \textbf{C}ontent. As such, we call our framework MIC. \newest{MIC is also able to represent how these affordances could impact each other by defining a notion of \emph{inter-affordance relationships}.} 
%\newwriting{As is the case with any ecosystem, these moderation-related affordances likely impact each other. To represent this, we have also included in MIC} a notion of \emph{inter-affordance relationships}. 

\newwriting{The MIC framework has key implications for moderation researchers \newest{and stakeholders.} More concretely, we argue that the advantages of using the MIC framework are three-fold: 
\begin{enumerate}
    \item The affordances and inter-affordance relationships in MIC provide a simple and explicit representation of potentially complex or subtle moderation ecosystems of social platforms. These components will also provide moderation researchers and community owners a convenient ``checklist'' to aid them in exploring and considering platforms to understand how moderation occurs on them. 
    \item MIC can be used to compare and contrast platforms' moderation ecosystems. Online community owners can use these comparisons to help decide which platforms would be more conducive for the moderation needs of their communities. Moderation researchers and platform designers can use these comparisons to uncover where platforms can adapt and learn from more established and better-studied platforms, as well as learn from the pitfalls these platforms have encountered. 
    \item MIC's representation of a platform's moderation ecosystem can be easily updated to reflect platform changes. Inter-affordance relationships can also be examined to catch potential moderation issues that new features could cause. This will allow moderation researchers \newest{and stakeholders} to update their understanding of platforms, and re-evaluate and potentially update moderation strategies and tools that might be impacted by platform changes. 
\end{enumerate} }

We will use MIC to analyze several social platforms through two case studies \newest{to exhibit these advantages}. Our first case study focuses on analyzing an individual platform using MIC, and shows how MIC can easily reflect platform changes \newest{(1)} as well as propagate such changes throughout the moderation ecosystem to account for new moderation challenges \newest{(3)}. In the second case study, we use MIC to systematically compare three platforms and use these MIC-based comparisons to propose potential moderation mechanisms that platforms can adapt from one another \newest{(2)}.

\section{Background}
\label{sec:relwork}
\newest{To better motivate our theoretical framework, we will first outline the relevant components of Grimmelmann's taxonomy \cite{grimm}, since our framework builds on top of these components. Then, we will briefly introduce each affordance while discussing the previous moderation work that informed our decision to include it in MIC.} 

%Grimmelmann's taxonomy Before detailing our framework, we introduce the platform affordances that we account for in MIC and review related work that motivated each of these affordances.
%First, we describe the high-level organization of these affordances, which was inspired by Grimmelmann's work \cite{grimm}. 

\subsection{\newest{Grimmelmann's Taxonomy of Moderation}}

Grimmelmann defines an online community using three features: the community's \emph{members}, the \emph{content} that is shared among the members, and the \emph{infrastructure} used to share it~\cite{grimm}. We use these features to \newest{delineate} the three main categories for affordances that are included in the MIC framework. \newest{We opt for this organization because each facet of this definition can impact moderators' ability to carry out Grimmelmann's four basic techniques for moderation \cite{grimm}}.  
%Now we discuss how each of these categories impacts the four basic techniques for moderation listed by Grimmelmann. 
\emph{Exclusion} is the act of excluding problematic or unwanted members from the community. Another closely related technique is \emph{pricing,} which controls the participation of community members by introducing barriers to entry. Both exclusion and pricing are mandated by the infrastructure and members of the community: infrastructure provides the tools for exclusion or pricing, while members are involved in using these tools. \emph{Organizing} is a technique that involves ``shaping the flow of content from authors.'' This technique is \newest{not only} impacted by the nature of content within the community, \newest{but also by the} infrastructure \newest{that provides certain members with such} ``shaping'' capabilities. Finally, \emph{norm-setting} involves the creation and articulation of community norms to establish acceptable types of behavior. Norm-setting is often accomplished by the other techniques, as well as by \newest{identifying and highlighting members of the community that exhibit these types of behaviors. In the following subsections, we will describe findings in moderation literature that speak to these claims.} 

%Next, we discuss each category of affordances included in our framework and review related work examining these affordances, with a particular emphasis on research related to moderation. %\newwriting{and audio-based social platforms, since our case studies in \Cref{sec:case-1,sec:case-2} will largely focus on them.}

\subsection{Member-related Affordances}
Through interviews with volunteer moderators of Discord servers,~\citet{jiang19} found that server owners create custom user roles to distinguish between various \emph{user types}.
The moderator role is a common facet of online communities and a role that is often assumed by volunteers on platforms relying on distributed moderation~\cite{seering2019moderator, wohn2019volunteer, jiang19, gilbert2020run}.
% Each role can be assigned specific permissions and thereby limit the access of certain users and to certain channels on the server.\emph{ User roles} and \emph{access} are thus closely related, and constitute two of three member-related components in our framework. 

The second member-related component in our framework is \emph{anonymity}. \citet{schlesinger17} 
% Schlesinger et al. 
studied how anonymity affects content on Yik Yak, a social media application that allowed users to make anonymous text posts that are grouped by location~\cite{schlesinger17} . In general, anonymity has been found to have both positive and negative effects on social interactions \cite{christopherson07}. 
Outside the context of online social spaces, anonymity was found to remove status markers that prevent members from participating in discussions on collaborative systems \cite{Hayne97,mcleod1997comprehensive,weisband1993overcoming}. 
Prior work examining the role anonymous voice-based interactions in online games found that in some cases anonymity was lost due to the nature of voice-based communication, and this caused some players to feel uncomfortable~\cite{wadley2015voice}. In fact, this loss of anonymity was deemed as one of the main reasons behind gamers abandoning the game being studied.

\subsection{Infrastructure-related Affordances}

One of the main infrastructural affordances we consider is a platform's \emph{organization}, i.e., how content and communities of the platform are situated. On Twitch, text-chats are associated to specific live streams, and live streams are separated by different Twitch channels; different streaming channels have \newest{different moderators and moderation strategies \cite{seering17}}. 
% Seering et al. \cite{seering17} studied how moderators encourage pro-social and discourage anti-social behavior in the text-chat of Twitch, a live video streaming platform. Discord also has more than one modality that the platform supports---text and audio channels.
% On Discord \cite{jiang19}, communities exist inside servers, and different servers have different sets of moderators and rules.
In certain cases, the lack of certain organizational structures within platforms might force \newest{communities to use a combination of two or more platforms to overcome these deficiencies.} This might lead to various \emph{inter-platform relationships}, which can be seen in prior work studying how moderators of Reddit communities use both Reddit and Discord to host their communities and the resulting challenges \newest{faced by multi-platform communities}~\cite{kiene19}.

Other integral parts of the infrastructure of ABSPs include the \emph{rules and guidelines} of platforms and the communities they host. Prior work has examined the rules that moderators of both Reddit and Discord outline for their communities, as well as guidelines specified by the platform itself~\cite{kiene19, jiang19}. 
Rules and guidelines, both community-defined and platform-specified, often describe the different roles members can play within the community (e.g., both Discord and Reddit have \newest{online manuals to help guide volunteer moderators.}
%pages dedicated to defining what the role of a moderator entails). 
\newest{Platform and community} rules and guidelines have also been shown to \newest{play a large role in} shaping community norms \cite{kiene16,cialdini1998social,triandis1994culture}. Platforms often have different \emph{badges and markers} \newest{to help users identify quality content or experienced users, and can be used to shape the behavior of users \cite{anderson13badges}. A common challenge encountered in video-based or voice-based communication systems is a lack of markers that provide cues to indicate when a user wishes to speak \cite{isaacs1994video,olson1995mix}).}

We designate a \emph{moderation mechanisms} \newest{affordance to represent a platforms' designated moderation tools or features, i.e. the infrastructure that a platform provides specifically for moderation.}
% Jiang et al. found that one of the biggest struggles when moderating Discord voice-channels is collecting evidence to prove that antisocial members were indeed engaging in antisocial behavior. The main reason behind this challenge was limited moderator time and availability---moderators were unable to be present in voice-channels at all times of use~\cite{jiang19}. 
Reddit has automated moderation tools, as well as an API that allows moderators to create moderation tools and bots to help human moderators to review large volumes of content. Discord has similar tools for moderators, some of which have been found to cause unprecedented moderation issues \cite{jiang19}. Prior work has explored how volunteer moderators employ a variety of mechanisms for moderating content, and moderation typically involves a large amount of time and effort to keep up with the massive amounts of content generated within social platforms~\cite{matias2019civic, kiene19}.
As a result, automated and human-machine collaboration tools have been developed to assist moderators on text-based platforms like Reddit \cite{jhaver2019human, chandrasekharan2019crossmod}. Video-hosting platforms like YouTube use algorithmic moderation \newest{to allow for a} larger moderation purview \newest{and to alleviate the labor of human moderators} \cite{gorwa2020algorithmic,roberts2016commercial}. Finally, platforms that enable \emph{monetization} may have unique moderation problems, since monetization has been found to lead to controversial behavior online to achieve virality \cite{bertaglia2021clout}, and algorithmic moderation tools can negatively impact users who rely on the monetization of their content \cite{oeldorf2021we}.

\subsection{Content-related Affordances}

\newwriting{Our framework considers the various \emph{modalities} platforms can support. As discussed in the previous subsections, the modality of content plays a role in how the content is viewed, organized, and moderated.} 
%Much of the communication that occurs in the audio-based social platforms discussed previously occurs in real-time. 
Voice-based communication systems and audio-based communication used in online gaming utilize real-time, or \emph{synchronous} audio \cite{ackerman97,tang09, wadley2015voice}. \citet{ackerman97} studied how users viewed and used Thunderwire, a collaborative audio-only real-time communication system modeled after telephone ``party lines'' of the late 19th century. \citet{wadley2015voice} studied real-time audio-communication in online multiplayer games and virtual worlds during game play. However, there are voice-based communities from India that use asynchronous audio for communication \cite{patel10,vashistha15}. 

\newest{An affordance closely related to synchronicity is} \emph{ephemerality}, as it is often, but not always, a consequence of synchronous or real-time content. Both communities studied by \citet{ackerman97} and \citet{wadley2015voice} used synchronous and ephemeral content. Prior work on ephemerality in social platforms has largely focused on ephemerality of text posts, links or images~\cite{schlesinger17, bernstein20114chan, Xu16}. \citet{jiang19} studied the challenges of moderating voice on Discord and found that the synchronicity and ephemerality of audio-based content created novel challenges for moderators.

Finally, social platforms can allow for certain \emph{access and restrictions} imposed on either viewing or creating content. In the past, subreddit moderators have purposely restricted access to their content as a way to express dissatisfaction with certain platform changes \cite{matias16dark}. Similarly, restrictions and access have been used to subdue antisocial behavior, though the efficacy of doing so is largely unclear \cite{tarvin2022youtube}. 
% \begin{itemize}
%     \item Also found some work on Spotify and how it has changed the relationship between musicians and their audience, and work that talks about music as a mechanism for communication.
%     \item I think I saw a recent IR paper that discusses how podcasts are ``advertised" on other platforms, so will snoop around that paper's related work to see if there's anything relevent there
%     \item Found a good amount of work on novel types of voice abuse that occur in VR/AR (includes some from Aaron's Discord paper)
%     \item Work about other audi-only social medias from years ago
%     \item work related to how the pandemic has altered users' relationship with technology
%     \item this is not an exhaustive list, just summarizing what I've looked into so far.
% \end{itemize}
%here
\section{MIC: A framework for representing the moderation ecosystem of social platforms}
\label{sec:feats}

\begin{figure}
    \centering
    \includegraphics[width =\textwidth]{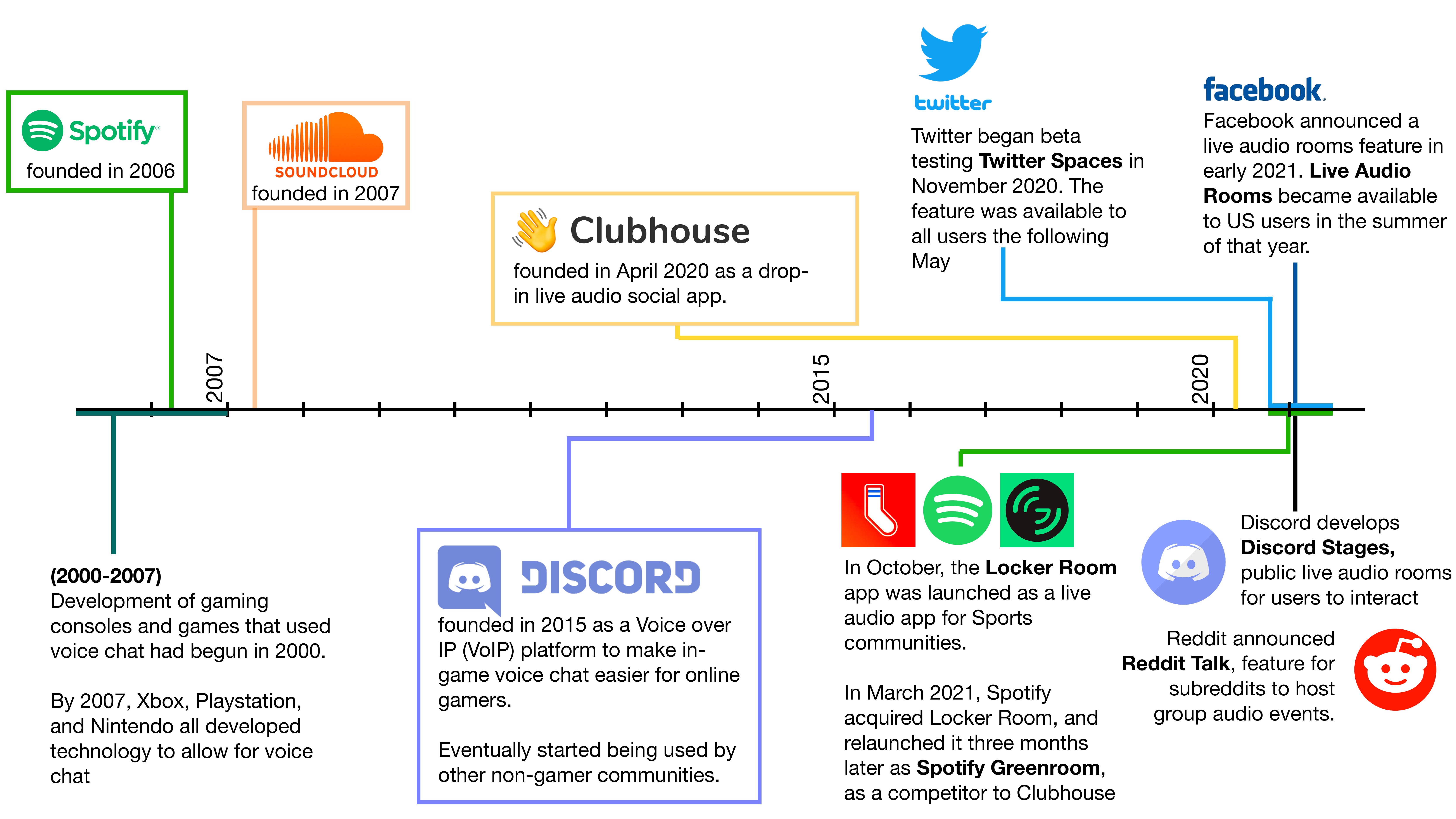}
    \caption{A timeline of popular audio-based technologies and social platform development. Clubhouse appears to mark the beginning of an audio-based ``boom'' in platform development\newest{, however platforms have utilized audio content since as early as 2000. We focus our attention on audio-based social platforms Spotify, Discord, Clubhouse, and occasionally Soundcloud, to showcase MIC's utility.} }
    \label{fig:tl-absps}
\end{figure}

\begin{figure}
    \centering
    \includegraphics[width=\textwidth]{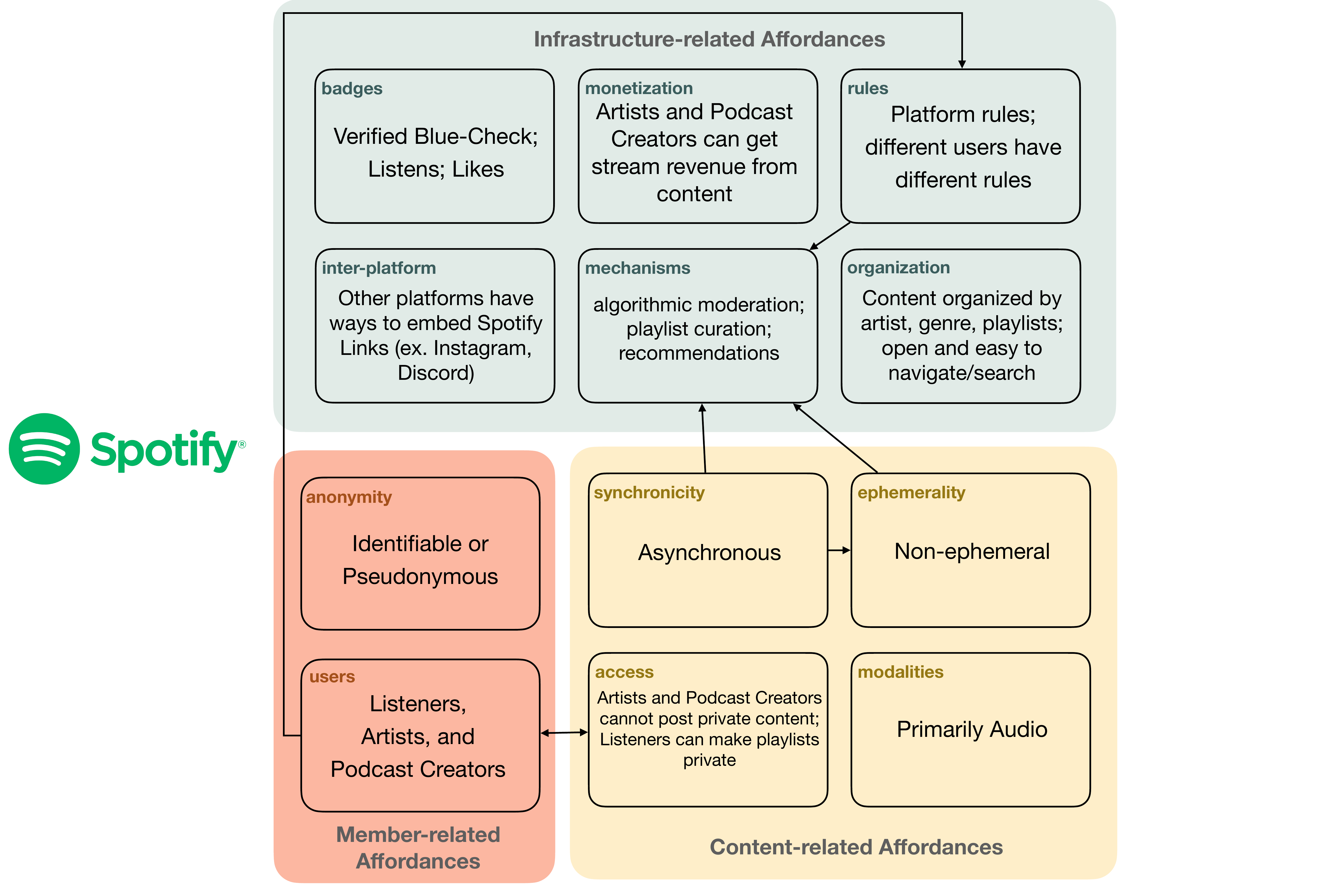}
    \caption{\newest{The MIC diagram for Spotify provides a graphical representation of its moderation ecosystem, and shows the inter-affordance relationships that occur using directed arrows. Moderation on Spotify is done primarily by the platform itself (\textbf{mechanisms}), as there is no moderator role available to users (\textbf{user}), and content can only be removed by the poster or the platform (\textbf{access}).}}
    \label{fig:spot}
\end{figure}

\begin{figure}
    \centering
    \includegraphics[width=\textwidth]{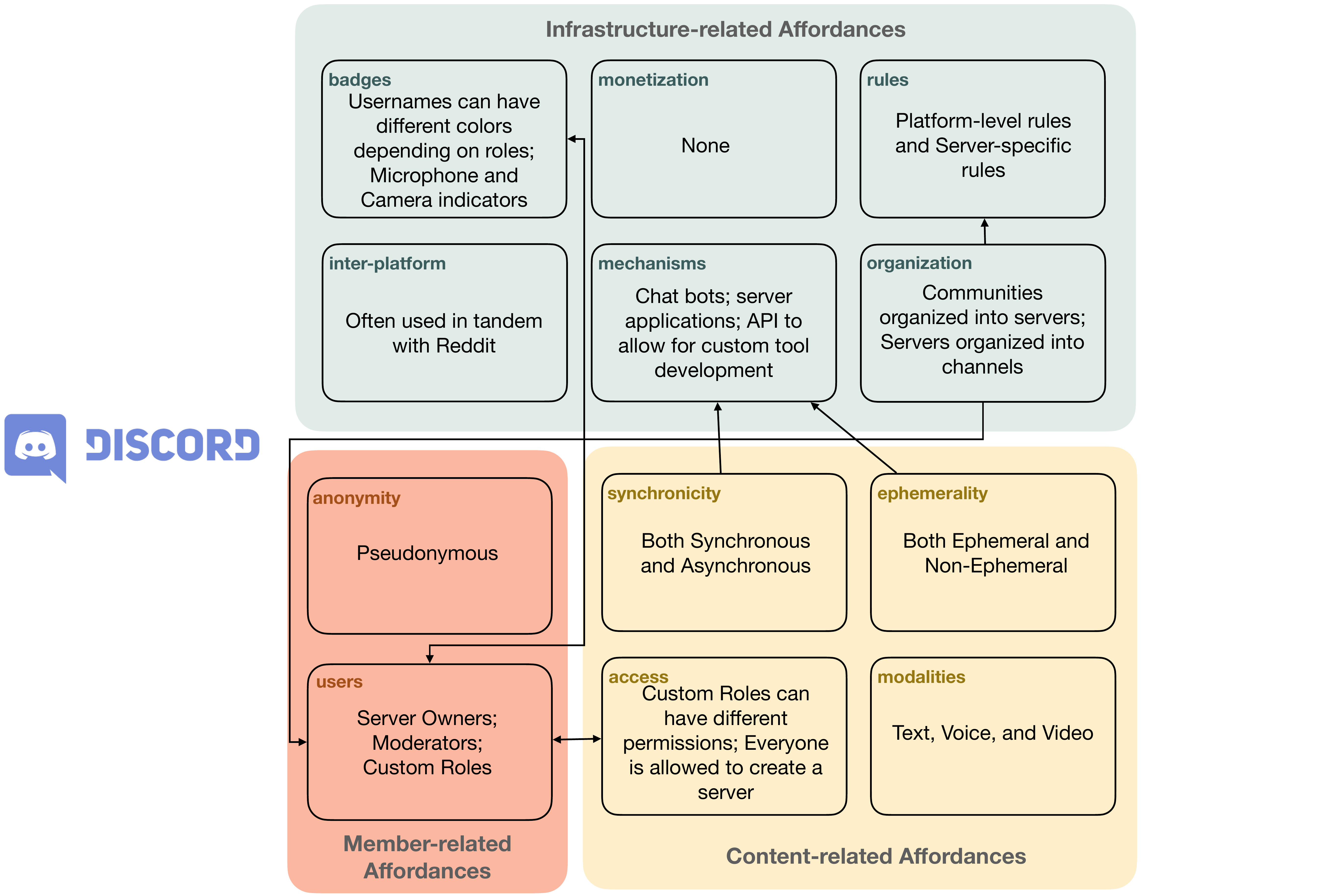}
    \caption{\newest{The MIC diagram for Discord. Moderation on Discord is done primarily by users who assume the moderator role in each server (\textbf{user}); moderators can use tools provided by the platform itself, or create their own tools to address their community's needs (\textbf{mechanisms}).}}
    \label{fig:disc}
\end{figure}

\newest{In order to formally define MIC and provide examples of its affordances and inter-affordance relationships, we will use three audio-based social platforms (Discord, Spotify, and occasionally Soundcloud) as working examples. These three platforms pre-date the audio-based social platform ``boom,'' which can be seen in the timeline shown in \Cref{fig:tl-absps}. For Discord and Spotify, we will construct \emph{MIC diagrams} (see \Cref{fig:spot,fig:disc}) to better highlight how MIC represents their moderation ecosystems and \newest{inter-affordance relationships that exist within them}. These MIC diagrams, as well as one we will construct for Clubhouse in \Cref{sec:case-1}, will be used to guide platform comparisons in \Cref{sec:case-2}. High-level descriptions of Spotify, Discord, and SoundCloud are provided below. 
}

% In this section, we formally define MIC through its components: platform \emph{affordances} and the \emph{relationships} between them. Affordances are properties of \newwriting{platforms that play a role in moderation.} We have identified three categories of affordances related to members, content and infrastructure. Together, these components can be used to create \emph{MIC diagrams} (see \Cref{fig:spot,fig:disc}) to \newwriting{highlight the moderation ecosystem of a platform.} 

% We will use the platforms Discord, Spotify, and Soundcloud as working examples to help us describe affordances and relationships. The affordance classifications and relationships from these examples were formed using participatory observations provided by the first author,

% as well as some prior work. We will also construct MIC diagrams for Spotify (\Cref{fig:spot}) and Discord (\Cref{fig:disc}) using the framework. 
%High-level descriptions of these platforms are provided below. 
% There are high-level descriptions of each platform below. We encourage the reader to further investigate each platform independently, as they are all free, popular, and prominent ABSPs. 

\begin{itemize}[label={}]
  \item \textbf{Spotify.} A audio-streaming service that hosts both music and podcasts. The main two types of Spotify users are listeners (those who use the service to stream content) and creators (those who use the service to upload content). Listeners are able to follow both creators and other listeners, and can view the latter's playlists and listening history. Creators must use other Spotify services, such as Spotify For Artists (for musicians) and Anchor (for podcasters).
  \item \textbf{SoundCloud.} A music-sharing website that allows all users to post audio (which consists of music, podcasts, random noises, etc). Users are able to comment on audio files and re-post others' audio posts on to their feed.
  \item \textbf{Discord.} A messaging platform that allow users to communicate via text, voice, or video. Discord's infrastructure is composed of ``servers,'' which can be thought of as landing pages for individual communities that use the platform. Servers can contain topic specific text-channels or voice/video channels. Server owners can create custom roles for server members, and can associate specific permissions for each role.
\end{itemize}

% TODO: new figures

\subsection{MIC Affordances}
\label{sec:feat-aff}

\newest{For each of MIC's twelve affordances, we}
%\newwriting{We present twelve affordances that can be used to represent social platforms in the MIC framework.}
%For each affordance, we 
\newwriting{provide a general description and highlight how they are supported by different platforms 
% identify variations of each affordance 
through our working examples}. 
We will continue to discuss how these affordances play a role in moderation on platforms. 

\subsubsection*{Modalities (\textbf{modalities})} \newest{\emph{Unimodal} platforms are centered around one primary type of modality, while \emph{multimodal} platforms use multiple types of modalities}. Discord is multimodal since servers contain text-\newwriting{and voice/video-} channels.  Spotify, on the other hand, is unimodal since audio is the primary type of content supported by the platform. \newwriting{The existence of multiple modalities} will affect moderation on the platform, since having more than one modality typically requires a broader set of policies and tools for moderation \cite{matias2019civic, jiang19, kiene19}.  

\subsubsection*{Access and Restrictions (\textbf{access})} Platforms often have various access and permission settings that allow or prohibit content from being posted, viewed, or removed. Many of these settings are accessible by the content creator, while some are limited to the platform. Discord allows server-owners and moderators to limit access to the server itself and to channels; the ability to use certain messaging features can also be limited by owners or moderators. Spotify only allows creators (musicians or podcasters) to publish content. Since Anchor is a free service for users who wish to become podcasters, there is no restrictions to post podcasts. However, users cannot publish music to Spotify directly--they must use a music distributor. Popular musicians are often signed to record companies or labels that will either act as or employ a distributor. Independent artists, those who do not have the backing of a record company, can use online music distribution services\footnote{DistroKid (https://distrokid.com/) is one such distribution service.} to publish music on Spotify. These services are never free, and therefore access to publishing music on Spotify is restricted. SoundCloud, on the other hand, allows all of its users to post audio-content, and only limits the amount of audio-content a free user can upload before requiring a paid SoundCloud Pro account. Spotify's and SoundCloud's access barriers are examples of Grimmelmann's pricing technique \cite{grimm}.

\subsubsection*{Monetization (\textbf{monetization})} Monetization \newest{refers to whether a platform provides infrastructure that enables users to generate revenue from their participation.}
%on platforms refers to whether content is being used to generate revenue for both the platform and content creator. 
\newest{Discord does not provide built-in mechanisms to enable monetization. Spotify monetizes all music, paying the artists according to how many times a song is streamed. Spotify also provides monetization tools through Anchor to allow podcast hosts to monetize their content as well.} 
%There is no content on Discord that can be monetized on the platform itself. Music and podcasts on Spotify are monetized, and creators receive profits based off of the number of streams their content receives. Soundcloud content is not monetized. 
Content that is being monetized may be more heavily moderated than content that is not; monetization may also incentivize creators to generate more content, which could lead to more moderation challenges.

\subsubsection*{Synchronicity (\textbf{synchronicity})} A platform's content is synchronous if it is being generated in real-time. Voice chats on Discord can only occur synchronously, whereas text-based conversations may occur asynchronously. Audio on Spotify is asynchronous. Synchronous content often creates challenges for moderators since not all moderators or moderation mechanisms can be present at the time the content is being created. Asynchronous content provides a larger window of opportunity for moderation mechanisms to detect and report abusive content or behavior.

\subsubsection*{Ephemerality (\textbf{ephemerality})} Ephemerality refers to whether or not the content can be accessed after it has been created and/or posted. On Discord, voice chats are ephemeral, since recording voice-channels can violate Discord's Terms of Service. On Spotify, audio is not ephemeral. Studies have shown that users behave differently when interactions are ephemeral and leave no record or trace~\cite{bernstein20114chan, schlesinger17}. Furthermore, when content is ephemeral, it becomes difficult for moderators to collect robust evidence to prove that anti-social behavior occurred to remove bad actors~\cite{jiang19}.

\subsubsection*{User \newwriting{Types} (\textbf{users})} Platforms may distinguish between types of users, and may even have designated types that allow users to act as moderators. Different user types are often associated with different permissions. On Discord, server owners and administrators can create custom roles for users, each with custom permission settings; one such role is typically assigned to ``moderators.'' On Spotify, only users with Spotify for Artist accounts are able to publish music. All users are able to create Anchor accounts to publish podcasts. Spotify has no designated ``Moderator''-like role assigned to users on the platform. 

% \subsubsection*{\newwriting{Permissions}} \remark{needs to be clear on how this differs from access for content} Types of permissions and restrictions users have for creating and consuming audio-based content on the ABSP. On Discord, though individual servers and channels might prohibit or limit the access provided to users, all users are able to create their own voice channels, thereby allowing free access to the creation and consumption of audio content.
% Spotify allows all users to consume audio, but only allows creators (musicians or podcasters) to create and publicly post audio-content. Since Anchor is a free service for users who wish to become podcasters, there is no restriction to post podcasts. However, users cannot publish music to Spotify directly--they must use a music distributor. Popular musicians are often signed to record companies or labels that will either act as or employ a distributor. Independent artists, those who do not have the backing of a record company, can use online music distribution services like DistroKid\footnote{https://distrokid.com/} to publish music on Spotify. These services are never free, and therefore access to publishing music on Spotify is restricted. SoundCloud, on the other hand, allows all of its users to post audio-content, and only limits the amount of audio-content a free user can upload before requiring a paid SoundCloud Pro account. The types of barriers to access on Spotify and SoundCloud are examples of the pricing moderation technique outlined by Grimmelmann \cite{grimm}.

\subsubsection*{Anonymity (\textbf{anonymity})} Users on platforms may be anonymous or use pseudonymous usernames to mask their identity. On Discord, users typically adopt usernames or handles that are custom and/or pseudonyms. 
Thus, users in voice-channels might not be not associated with any actual means of identification. On Spotify, listeners can, and often do, create account usernames with their actual identity (typically by linking Spotify to other social media accounts, such as Facebook). However, some users do adopt custom usernames that obscure their identity. Creators may publish audio-content under stage names or aliases. Anonymity has been found to both enable and discourage negative behavior in online social spaces \cite{Hayne97}, and anonymity appears to break down when using voice-based communication \cite{wadley2015voice}.

\subsubsection*{\newwriting{Organization (\textbf{organization})}} The organization of a platform refers to the way in which content and communities are organized, situated, and discovered on the platform. A platforms' organization impacts users' and moderators' ability to locate content and members of interest. Discord is organized into servers, and each server has various channels in which community members interact and share content. Users can use Discord's Server Discovery feature or Explore page to look for popular public servers to join, or create their own public or private servers. Not all large servers are necessarily public or searchable using Discord's Server Discovery. The vast majority of audio-content on Spotify is indexed and publicly available to every user of the service. Typically, audio on Spotify is organized by artist, genre, podcast, or in user- or algorithmically-curated playlists (some of which are private). Users can search and discover all public audio-content via search or using Spotify's various discovery and recommendation mechanisms. 

% Discord are much more \emph{confined}, since they are situated inside designated servers, most of which are not easily discoverable or indexed properly.  In this way, Spotify is more \emph{free range} compared to Discord. 
% %Structured ABSPs may allow for better moderation tools. For instance, the server structure of Discord allows moderators to create community-specific rules and tools to help with moderation \cite{kiene19}.
% % We should be Organization in the context of CMI-F should 

\subsubsection*{Rules and Guidelines (\textbf{rules})} Most platforms utilize some combination platform-wide terms of service (TOS) and community-specific guidelines to govern user behavior. These terms and guidelines establish high-level rules that all users are expected to abide by. In addition to community guidelines and TOS, Discord also has platform-level rules that clearly define the roles of moderators on servers. At the community-level, Discord servers can publish their own set of rules and guidelines that are typically more tailored to the type of community the server hosts. Spotify has separate guidelines and TOS for listeners and content creators who use Spotify for Artists and Anchor. The rules and guidelines help establish a baseline for both platform-wide and community-specific norms and conditions for exclusion (e.g., suspensions or bans~\cite{chandrasekharan2017you}). Rules and guidelines play a key role in moderation, as seen in Grimmelmann's work---\textit{norm-setting} and \textit{exclusion} make up two of the four common techniques for moderation~\cite{grimm}.

\subsubsection*{\newwriting{Badges} and Markers (\textbf{badges})} Badges and markers refer to the various types of visual cues or indicators that could be applied to users and content.
On Discord, different user types can have different colors associated with them. For example, if a ``moderator'' role is associated with the color red on a Discord server, we know that a user's handle (i.e., username) appearing in red indicates that the user is a moderator.
Such markers help other members identify the official moderators of a server, and depending on what other roles the server defines, could help identify different types of users. Discord also provides indicators that show whether participants of a voice call have their microphone muted or their video on; this information can be seen without having to actually join the voice-call. On Spotify, artists with a ``verified'' blue-check mark on their profile which indicates that the identity of the owner of the artist page has been officially verified by Spotify. This signal indicates to users that the content posted on this artist's page is coming from an official source. Spotify also displays the number of times a song has been listened to and the number of users who have liked a playlist. Such badges and markers help in moderation since they provide users and moderators with additional cues to determine whether certain users or content are safe to engage with. 

\subsubsection*{Inter-Platform \newwriting{Relationships} (\textbf{inter-platform})} 
The way users of one social platform utilize other platforms is an aspect that is not often highlighted when discussing moderation on social platforms in general.
% The way ABSPs and other platforms utilize other platforms is an often overlooked aspect when discussing moderation on SNSs in general. 
Discord servers are known to be used alongside other platforms (such as Reddit \cite{kiene19}), but are also commonly used alone. Discord users will occasionally use other, more free-range platforms such as Twitter and Reddit to discover and advertise private servers. Spotify, on the other hand, is often used by other platforms to embed music. For instance, Instagram users can add music directly from Spotify to their story posts, or link to their Spotify playlists. As more SNSs become available, it will be more commonplace for online communities to use more than one platform. This affects moderation since bad actors can harass users over multiple platforms, making moderation more difficult \cite{jhaver18}.

\subsubsection*{Moderation \newwriting{Mechanisms} (\textbf{mechanisms})} \newwriting{The moderation mechanisms of a platform refer to its built-in moderation tools and procedures. Discord allows users to use and create chat bots and tools to moderate text-channels. Discord also has a guide for moderators. However, not all interactions in a voice-channel can be moderated unless a moderator is present in the voice-channel every time there is activity or the voice-channels are being recorded. Discord has bots that enable recording, but depending on where users reside, consent must be granted in order for recording to be allowed.} On Spotify, all audio content \emph{can} be moderated by the platform itself, since audio must be first uploaded to the platform and processed before it is hosted publicly. Spotify has mechanisms for algorithmic content moderation; this is the case with moderating copyright-abiding content \cite{brovig2021remix}. and the existence of such mechanisms leads us to believe that all audio-content is moderated in some way. Limited moderation mechanisms allow abusive and antisocial behavior to go unchecked on social platforms. 

\subsection{Relationships Between Affordances}
\label{sec:feat-rel}

Though we have defined a set of disjoint affordances, \newest{affordances can have relationships with each other.} 
%these affordances will often be linked to each other in the larger platform ecosystem. 
For instance, in both Spotify and Discord, access is linked to user roles, since different types of roles constitute different types of access. Highlighting these inter-affordance relationships will illuminate how potential modifications to one affordance could impact \newest{the moderation ecosystem at large}. Moreover, if a specific affordance has been identified as a contributor to moderation challenges, we can use inter-affordance relationships to identify other, less apparent affordances that also contribute to these challenges. 

\newwriting{Formally, we define an inter-affordance relationship from affordance $A$ to affordance $B$ if modifying affordance $A$ impacts or changes the status of affordance $B$. For example, the asynchronous nature of content on Spotify (\textbf{synchronicity}) enables its non-ephemerality (\textbf{ephemerality}); indeed, if Spotify introduced synchronous content, then the ephemerality of certain content might change\footnote{In fact, Spotify Greenroom has synchronous and ephemeral content, and gives users to option to record live audio rooms to upload to Spotify.}. On Discord, the \textbf{ephemerality} and \textbf{synchronicity} of the voice interactions in voice-channels affect the moderation mechanisms that are available on the platform. In our MIC diagrams, these relationships are shown as directed arrows between affordances. A bi-directional arrow is used to indicate when a relationship exists in both ``directions.'' For example, \textbf{user} types on both Spotify and Discord are tied to types of \textbf{access} and permissions. These relationships in a platform will likely change over time as the platform itself is updated.}

%To further reinforce our notion of inter-affordance relationships, we list more of the relationships that exist among the affordances of Spotify and Discord. 
\newest{Other inter-affordance relationships in Spotify and Discord's moderation ecosystems are as follows:}
The non-ephemeral (\textbf{ephemerality}) and asynchronous (\textbf{synchronicity}) nature of content on Spotify affects the platforms' moderation \textbf{mechanisms}. Similarly, the moderation \textbf{mechanisms} are enabled by Spotify's user agreement, which explicitly states that the platform is allowed to remove or edit any content that is uploaded if it violates community guidelines (\textbf{rules}). On Discord, user types are often \newest{are often customized to each server,} thus the \textbf{organization} of Discord has an affect on \textbf{user} types.

\section{MIC as a tool for analyzing individual platforms}
\label{sec:case-1}

In this section, we will demonstrate how MIC can be used to represent and subsequently update our understanding of a particular platform's moderation ecosystem. We will use MIC to analyze the Clubhouse app, which has been rapidly evolving since its release in 2020, at two different points in time. 

\begin{figure}
    \centering
    \includegraphics[width=\textwidth]{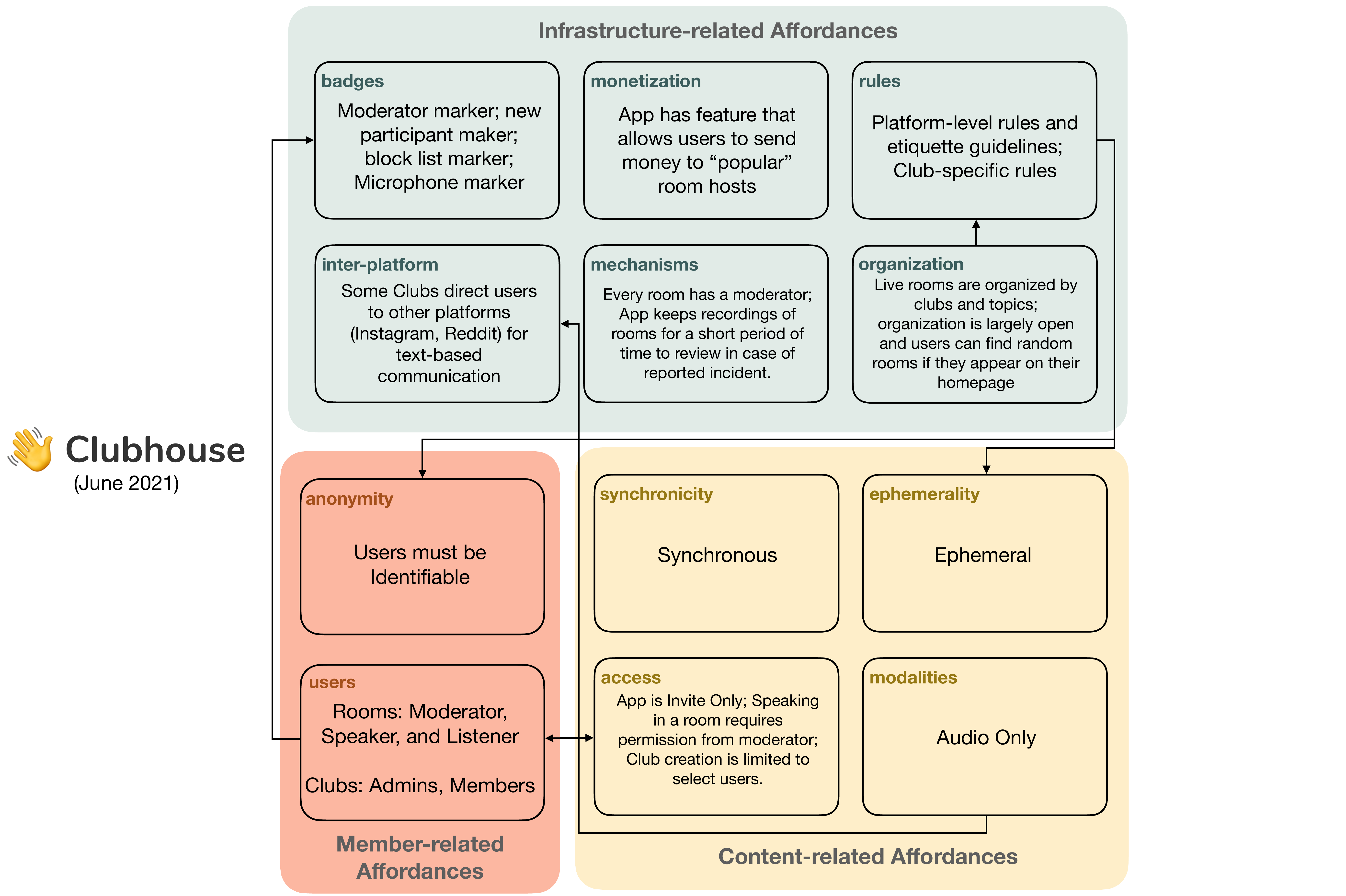}
    \caption{MIC diagram for Clubhouse as of June 2021. \newest{At this point, Clubhouse users have no way of interacting with each other apart from using synchronous and ephemeral audio spaces (\textbf{modalities}); this led to many Clubhouse users using other platforms to engage in asynchronous text-based communication (\textbf{inter-platform}).}}
    \label{fig:ch-21}
\end{figure}

First, we will describe Clubhouse \newest{and its affordances} as of June of 2021, \newest{and in doing so will construct its MIC diagram} (\Cref{fig:ch-21}). \newest{Then, we describe the changes made to Clubhouse between June of 2021 and January of 2022 that could potentially impact moderation, updating the MIC diagram accordingly (\Cref{fig:ch-22}).}
%We then describe the state of Clubhouse as of the time of writing this manuscript, and accordingly update the MIC diagram and discuss how these changes could effect potential moderation challenges and strategies (\Cref{fig:ch-22}). 
Finally, we will discuss how using MIC allows us to reason about moderation strategies and challenges that exist on Clubhouse in a more efficient and systematic way, and what insights MIC provides that may otherwise be overlooked.

\subsection{\newest{Exploring} Clubhouse Using MIC}

\subsubsection*{Clubhouse (as of June 2021)} Clubhouse is an invite-only, so new users had to be invited to the app using their phone number (\textbf{access}). The platform's community guidelines require users to use their real names (\textbf{anonymity}). Clubhouse users can only communicate with one another using audio in public or private voice rooms (\textbf{modalities}). Clubhouse is organized into topic-specific pages and groups called ``clubs'' (\textbf{organization}); only ``the most active members of the Clubhouse Community'' can create clubs (\textbf{access}). Each such page and club is made up of synchronous and ephemeral voice rooms (\textbf{synchronicity, ephemerality}). Every club has designated admins that have the ability to edit the club settings, name, and manage members (\textbf{users}). Public voice rooms can be accessed by any user on the app, regardless of their membership in its associated club or interest in the room's subject (\textbf{access}). Private rooms can only be joined by the followers of the room host or the members of the room's associated club (if it exists) (\textbf{access}).  All participants of rooms are required to follow Clubhouse's Community Guidelines \cite{ch-cg} (\textbf{rules}). However, established clubs can publish a list of club-specific rules that can be applied to participants of rooms hosted by the club (\textbf{rules}). 

Users can have one of three roles in a room on Clubhouse (\textbf{users}). The moderator role (denoted by a green star symbol) is given to the user who creates the room. This user has the ability to end the room, invite users to the stage to speak, mute speakers, and assign other users to be moderators as well. This means that every active room (i.e., every instance that audio-content is generated on the app) has a ``moderator'' present (\textbf{mechanisms}). All other users that enter the room start out as listeners, and do not have the ability to speak in this role---they cannot unmute their microphone. As a listener, users can press the ``raise hand'' button and ask to be a speaker. If a moderator accepts a listener's request to speak, that listener gets moved up to the ``stage'' where they now have the role of speaker. As a speaker, they can unmute their own microphone and be heard by everyone else in the room (\textbf{access}). 

All speakers inside a room have a marker to show whether their microphone is muted or not. Speakers often click this marker on and off to indicate that they wish to speak next. When users enter a room, they have a celebratory emoji by their icon and name to indicate that they are new to the room (\textbf{badges}). Clubhouse also has a monetization feature that lets users send money to other Clubhouse users through their profile page (\textbf{monetization}). Clubhouse uses a block-list icon to indicate to a user when an account has been by many people in their circle (\textbf{mechanisms, badges}). 

\newest{Text-based commentary or discourse pertaining to the interactions that occur on Clubhouse} often happens on other platforms. One such platform that is heavily used \newest{to discuss particular events or incidents that happen in Clubhouse rooms is Twitter}. Users will often talk about what they are experiencing on Clubhouse on Twitter, and Clubhouse users will often link to their Twitter profiles on their Clubhouse profile. There are also subreddits dedicated to talking about Clubhouse (i.e., r/Clubhouse). These other platforms are also used to announce and publicize rooms or clubs and invite new users to Clubhouse (\textbf{inter-platform}).

% \subsection{MIC for Clubhouse in January of 2022}

\begin{figure}
    \centering
    \includegraphics[width=\textwidth]{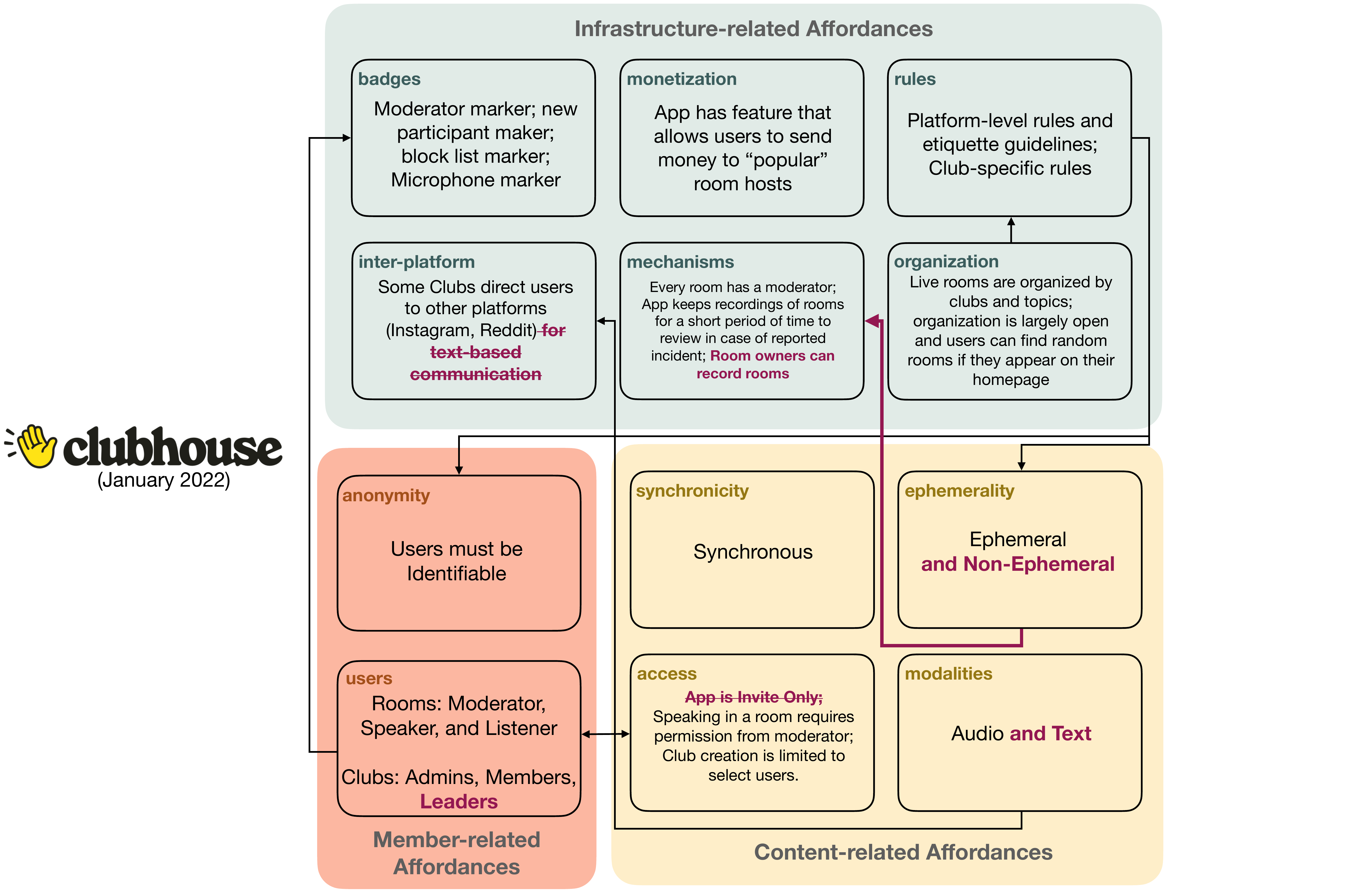}
    \caption{\newest{Updated MIC diagram for Clubhouse which reflects platform changes made between June 2021 and January 2022 (changes are shown in violet). Some major changes include the addition of a text-based direct-messaging feature (\textbf{modalities}) and the ability to keep recordings of live rooms on the app (\textbf{ephemerality}). These new additions lead to some changes to the inter-affordance relationships on the platform.}} 
    \label{fig:ch-22}
\end{figure}

\subsubsection*{\newest{Moderation-Related Updates to Clubhouse}} \newer{Between June of 2021 and January of 2022, Clubhouse released close to 20 updates to their iOS app \cite{iosRelease}. These releases included changes to the app's appearance, updates to the app's terms of service and privacy policy, as well as the addition of multiple new features. Using MIC, we identified which of these updates to investigate further to understand moderation on Clubhouse. The relevant changes are as follows: }\newwriting{Clubhouse is no longer invite-only, i.e., anyone with a smartphone is allowed to make an account and join the Clubhouse community (\textbf{access}). The platform also added a direct-messaging feature that lets users send text-messages to other users and create group chats (\textbf{modalities}). Clubs can now assign users a ``Leader'' role that gives them the ability start and schedule rooms in a club, but does not allow them to alter the club settings or add/remove members (\textbf{users}). By far the largest change to Clubhouse is that it introduced non-ephemeral content, i.e., live audio rooms can be recorded for users to listen to later (\textbf{ephemerality}). Additionally, Clubhouse added an option that lets users block inappropriate or NSFW voice rooms from their feed (\textbf{mechanisms}).}

\subsection{Insights into Moderation on Clubhouse} 

The observed affordances and relationships in MIC give us several insights into moderation on Clubhouse. First, the existence of the moderator role in every live audio room indicates that moderation on Clubhouse is done primarily by users as opposed to by the platform itself (\textbf{mechanisms}).  The platform's requirement of using identifiable information (\textbf{rules}) will impact the types of interactions that users have on the platform, \newest{and might impact the nature of antisocial behavior that occurs on the platform.} 
%discourage  hopefully reducing the frequency of antisocial behavior.} 
The organization of live audio rooms on Clubhouse will make it easy for users to discover new rooms and interact with new people (\textbf{organization}). However, this organization also lets users to abruptly leave rooms, which may make it difficult for room hosts and moderators to report disruptive or antisocial users. \newest{Clubhouse's record feature might allow room hosts to maintain records of users that engaged in disruptive behavior, as well as evidence of such behavior (\textbf{ephemerality, synchronicity}).}
%These records could later be used locate and reprimand problematic users}.

Before Clubhouse added a text-based chat feature, users had to utilize other social platforms if they wanted to send asynchronous, text-based messages to other users. This \newest{led to instances where abusive users used} several other platforms to harass individuals they initially encountered on Clubhouse \cite{lorenz_2021}. This type of behavior could amplify the amount of harassment a potential victim receives. Thus, the introduction of text-based messages (\textbf{modalities}) could likely reduced the \newest{reliance on these} inter-platform relationships, making Clubhouse, \newest{and anti-social behavior that occurs on Clubhouse,} more self-contained. This could potentially \newest{prevent the amplification of} harassment that victims of antisocial users get. Finally, since Clubhouse is no longer invite-only (\textbf{access}), the user base of Clubhouse has undoubtedly expanded. This means more users, and more communities, will start using Clubhouse, likely resulting in a large influx of user and incident reports, thereby posing newer challenges to the platform.

\section{MIC as a tool for cross-platform moderation analysis}
\label{sec:case-2}

So far, we have created MIC diagrams for three platforms, all of which are centered around audio. As discussed in the introduction, these audio platforms have many similarities and differences that could impact how moderation is accomplished. In this section, we will compare and contrast the platforms via the MIC framework. Then, we will use these comparisons to generate ideas for new moderation interventions.

\subsection{Similarities and Differences between Discord, Spotify, and Clubhouse}

\newest{We begin by pointing out} the obvious similarities and differences between the three platforms that can be determined without using MIC. First, Discord and Clubhouse both offer live audio features, whereas Spotify itself does not. Spotify also does not offer users a way to direct-message other users, while Discord and Clubhouse both have such features. In fact, Spotify users have no means to interact with one another on the platform apart from using posted audio, which is not the case on Discord or Clubhouse. In general, Spotify is used for listening to Music and Podcasts; Clubhouse is used for listening to and participating in live audio rooms; Discord is used to host communities and let community members interact with each other over text, voice, and video.

\newest{Affordance-based comparisons between Discord, Spotify, and Clubhouse are shown \Cref{tab:mic-comp}. Similarities and differences between inter-affordance relationships of the platforms can be seen by comparing the edges of the MIC diagrams found in \Cref{fig:disc,fig:spot,fig:ch-22}.}

% \usepackage{graphicx}
\begin{table}[]
\centering
\resizebox{\textwidth}{!}{%
\begin{tabular}{|c|c|c|c|}
\hline
\textbf{\begin{tabular}[c]{@{}c@{}}~\\ ~\end{tabular}} &
  \textbf{Discord} &
  \textbf{Spotify} &
  \textbf{Clubhouse} \\ \hline
\textbf{\begin{tabular}[c]{@{}c@{}}~\\ modalities\\ ~\end{tabular}} &
  \begin{tabular}[c]{@{}c@{}}~\\ Text, audio, \\ and video\\ ~\end{tabular} &
  \begin{tabular}[c]{@{}c@{}}~\\ Primarily audio\\ \\ ~\end{tabular} &
  \begin{tabular}[c]{@{}c@{}}~\\ Primarily audio\\ ~\end{tabular} \\ \hline
\textbf{\begin{tabular}[c]{@{}c@{}}~\\ access\\ ~\end{tabular}} &
  \begin{tabular}[c]{@{}c@{}}~\\ All users can \\ generate content\\ ~\end{tabular} &
  \begin{tabular}[c]{@{}c@{}}~\\ Not all users can\\ generate content\\ ~\end{tabular} &
  \begin{tabular}[c]{@{}c@{}}~\\ All users can \\ generate content\\ ~\end{tabular} \\ \hline
\textbf{\begin{tabular}[c]{@{}c@{}}~\\ monetization\\ ~\end{tabular}} &
  \begin{tabular}[c]{@{}c@{}}~\\ No monetization \\ tools for users\\ ~\end{tabular} &
  \begin{tabular}[c]{@{}c@{}}~\\ Content is monetized \\ (through ads and stream \\ count)\\ ~\end{tabular} &
  \begin{tabular}[c]{@{}c@{}}~\\ Users can be monetized\\ (other users can donate \\ to user profiles)\\ ~\end{tabular} \\ \hline
\textbf{\begin{tabular}[c]{@{}c@{}}~\\ synchronicity\\ ~\end{tabular}} &
  \begin{tabular}[c]{@{}c@{}}~\\ Synchronous (audio/video)\\ and Asynchronous (text)\\ ~\end{tabular} &
  \begin{tabular}[c]{@{}c@{}}~\\ Asynchronous (audio)\\ ~\end{tabular} &
  \begin{tabular}[c]{@{}c@{}}~\\ Synchronous (audio)\\ ~\end{tabular} \\ \hline
\textbf{\begin{tabular}[c]{@{}c@{}}~\\ ephemerality\\ ~\end{tabular}} &
  \begin{tabular}[c]{@{}c@{}}~\\ Ephemeral (audio/video)\\ and Non-ephemeral (text)\\ ~\end{tabular} &
  \begin{tabular}[c]{@{}c@{}}~\\ Non-ephemeral (audio)\\ ~\end{tabular} &
  \begin{tabular}[c]{@{}c@{}}~\\ Ephemeral (audio)\\ and Non-ephemeral (audio/text)\\ ~\end{tabular} \\ \hline
\textbf{\begin{tabular}[c]{@{}c@{}}~\\ organization\\ ~\end{tabular}} &
  \begin{tabular}[c]{@{}c@{}}~\\ All communities organized \\ into/within servers\\ \\ Users can search for public\\ servers using keyword search\\ ~\end{tabular} &
  \begin{tabular}[c]{@{}c@{}}~\\ Open organization; content\\ organized by genre, artist, \\ playlists etc.\\ ~\end{tabular} &
  \begin{tabular}[c]{@{}c@{}}~\\ Some communities\\ organized into/within clubs\\ \\ Public clubs and rooms \\ categorized by topic\\ ~\end{tabular} \\ \hline
\textbf{\begin{tabular}[c]{@{}c@{}}~\\ users\\ ~\end{tabular}} &
  \begin{tabular}[c]{@{}c@{}}~\\ Custom user roles can be\\ created for each server\\ ~\end{tabular} &
  \begin{tabular}[c]{@{}c@{}}~\\ Fixed, platform-wide user\\ roles (associated with a user's\\ account)\\ ~\end{tabular} &
  \begin{tabular}[c]{@{}c@{}}~\\ Fixed user roles that can change\\ within Clubs and rooms\\ ~\end{tabular} \\ \hline
\textbf{\begin{tabular}[c]{@{}c@{}}~\\ anonymity\\ ~\end{tabular}} &
  \begin{tabular}[c]{@{}c@{}}~\\ Pseudonymity allowed\\ ~\end{tabular} &
  \begin{tabular}[c]{@{}c@{}}~\\ Pseudonymity allowed\\ ~\end{tabular} &
  \begin{tabular}[c]{@{}c@{}}~\\ Users must be Identifiable\\ ~\end{tabular} \\ \hline
\textbf{\begin{tabular}[c]{@{}c@{}}~\\ rules\\ ~\end{tabular}} &
  \begin{tabular}[c]{@{}c@{}}~\\ Platform-wide and \\ server-specific rules\\ ~\end{tabular} &
  \begin{tabular}[c]{@{}c@{}}~\\ Platform-wide rules\\ ~\end{tabular} &
  \begin{tabular}[c]{@{}c@{}}~\\ Platform-wide and \\ club-specific rules\\ ~\end{tabular} \\ \hline
\textbf{\begin{tabular}[c]{@{}c@{}}~\\ badges\\ ~\end{tabular}} &
  \begin{tabular}[c]{@{}c@{}}~\\ Custom roles can\\ have custom visual markers\\ ~\end{tabular} &
  \begin{tabular}[c]{@{}c@{}}~\\ Verified Artist accounts have\\ ``Blue Check''\\ ~\end{tabular} &
  \begin{tabular}[c]{@{}c@{}}~\\ Users in rooms have badges \\ associated with their roles\\ ~\end{tabular} \\ \hline
\textbf{\begin{tabular}[c]{@{}c@{}}~\\ inter-platform\\ ~\end{tabular}} &
  \begin{tabular}[c]{@{}c@{}}~\\ Can host communities \\ independently\\ \\ Often used in tandem with\\ other platforms\\ ~\end{tabular} &
  \begin{tabular}[c]{@{}c@{}}~\\ Cannot host communities\\ independently\\ \\ Integrated into other platforms\\ ~\end{tabular} &
  \begin{tabular}[c]{@{}c@{}}~\\ Can host communities \\ independently\\ \\ Often used in tandem with\\ other platforms\\ ~\end{tabular} \\ \hline
\textbf{\begin{tabular}[c]{@{}c@{}}~\\ mechanisms\\ ~\end{tabular}} &
  \begin{tabular}[c]{@{}c@{}}~\\ Automated tools for\\ user-driven moderation\\ ~\end{tabular} &
  \begin{tabular}[c]{@{}c@{}}~\\ Automated tools for \\ platform-driven moderation\\ (through content flagging, \\ curation, and recommendation)\\ ~\end{tabular} &
  \begin{tabular}[c]{@{}c@{}}~\\ Platform-driven moderation \\ involves recording all audio rooms\\ \\ No automated tools for \\ user-driven moderation\\ ~\end{tabular} \\ \hline
\end{tabular}%
}
\caption{\newest{MIC-guided comparisons between the affordances of moderation ecosystems on Discord, Spotify, and Clubhouse. These comparisons will guide us in generating ideas for new moderation interventions for each platform.}}
\label{tab:mic-comp}
\end{table}

\subsection{Adapting and Proposing Moderation Mechanisms using MIC Comparisons}
\label{sec:case-2-mm}

\subsubsection*{Spotify and Clubhouse} One challenge we noticed while using Clubhouse to conduct the previous case study (\Cref{sec:case-1}) is that it was difficult to identify live rooms that are of interest that appear on the app's home page. Furthermore, some live rooms dealt with sensitive topics, such as sexual assault. Such rooms should likely not be \newest{recommended or} shown to users who are insensitive to certain topics, since their participation in the room would have negative impacts on the members of such a space. In general, it seems difficult for both listeners to find relevant and interesting rooms on Clubhouse and room hosts to find interested listeners and participants. To begin addressing this potential challenge, one can use \newest{MIC-based comparisons} to observe that Clubhouse has a similar open \textbf{organization} to Spotify. In particular, the room topic categories that users can browse on Clubhouse are reminiscent of the various categories users can use to browse content on Spotify. Likewise, both platforms host non-ephemeral content (\textbf{ephemerality}).

One of Spotify's major services is its recommendation system for music and podcast discovery. Not only does this service aim to show users content that they would be inclined to listen to, but also for creators to discover new listeners.\footnote{Both Anchor.fm and Spotify For Artists have tools for musicians to help them expand their audience.} One way in which Spotify does this is by curating playlists. These playlists can be broadly defined, containing music from a genre, or from a specific musical artist. Many of these playlists are manually curated, and artists can submit music for consideration to be added to these curated playlists. 

Given Clubhouse and Spotify's organizational similarity, and the existence of non-ephemeral content, we could propose a moderation mechanisms for Clubhouse that involves adopting a similar type of recommendation-via-curation mechanism like Spotify, and manually curate endorsed playlists of \newer{recordings of quality room recordings. We could even try to extend this idea to ephemeral content, i.e. playlist-type hubs of clubs or upcoming scheduled rooms that are hosted by trusted or experience users. This could start to help clubs and rooms find relevant audiences, and could also help users find and build communities in a more strategic way, while limiting the number of potential bad actors that try to engage.}

\subsubsection*{Discord and Clubhouse} \newwriting{MIC also showed us that Clubhouse and Discord are very similar across many different affordances. Discord has been studied in the context of moderation research \cite{kiene16,jiang19}, and researchers have found that moderating voice channels on Discord is a challenging feat. This is largely due to the fact that moderators in Discord servers find it difficult to monitor events and collect evidence of bad behavior in voice channels \cite{jiang19}. Clubhouse, like Discord, has a moderator role for users (\textbf{users}); however, on Clubhouse, every active room must have a moderator present. A feature, or moderation \textbf{mechanism}, that Discord could ``borrow'' from Clubhouse to help moderators handle voice-channels is a way to enable moderators to schedule when voice-channels can be made active . This way, moderators can ensure that they are present in public voice channels. Discord moderators can already limit when voice channels are open, but scheduling such time (similar to how live rooms are scheduled in Clubhouse clubs by Leaders and Admins) can make this easier to do.}
 
\newwriting{Another change Discord could make is adopt Clubhouse's policy of keeping recordings of voice-rooms for a short period of time in order to address or investigate any reports (\textbf{rules}). It might be the case that some Discord servers have such a policy for their server; creating a platform-wide policy would be a more robust measure to discourage harmful behavior in such spaces. However, the pseudonymous nature of Discord (\textbf{anonymity}) might make such a policy not only difficult to implement, but also off-putting to Discord's user base \newest{(a large portion of which is comprised of gaming communities, where users have been shown to prefer some degree of anonymity and privacy \cite{wadley2015voice})}. \newest{Clubhouse's recording policy, when introduced, did not appear to drive away its user base. Though it is unclear exactly why this was the case, it could be because} every user on the app has already agreed \newest{(in order to satisfy Clubhouse's Terms of Service)} to be identifiable, and thus users have already agreed to forfeit some of their privacy. Clubhouse can adapt some moderation \textbf{mechanisms} from Discord as well. In particular, Clubhouse could develop an API or a collection of chat bots or tools that help to moderate text conversations. Such tools could also be developed for room moderators to help them keep track of members of a room, flag certain users, handle requests to speak, or manage music streams, as is the case with certain Discord bots \cite{jiang19}. It might be the case that different types of rooms or clubs want or need different types of tools, thus the customizability of Discord's moderation tools and API could be useful for Clubhouse users.}

\section{Discussion}

For CSCW theory, our framework provides a new analytic lens to identify, understand, and compare the various components \newwriting{of a social platform's moderation ecosystem.} MIC allows \newest{moderation researchers and stakeholders} to efficiently and comprehensively navigate moderation-specific aspects of social platforms. \newer{The various insights MIC led us to can be used to develop research questions that moderation researchers can use to further investigate new and dynamic platforms like Clubhouse and motivate future studies. Likewise, platform designers and moderators themselves can use these insights to preemptively \newest{infer} potential moderation challenges that might arise, and can prepare for them by designing new tools, features, or guidelines.} \newer{Comparing moderation ecosystems across platforms using MIC can allow stakeholders to adopt successful moderation mechanisms from one another without overlooking subtle but potentially significant differences. We now discuss further implications, potential limitations, and extensions of MIC.}

\subsection{Implications and Advantages of Using MIC}

\subsubsection*{Efficient Navigation of New Platforms} \newwriting{Platforms often offer a plethora of features which can make it difficult to discern which features are relevant for moderation. MIC allows us to systematically pinpoint the facets of a platform's design and affordances that are relevant. In our case studies, we used MIC to determine relevant features to examine their role in effecting moderation on different platforms. For instance, Clubhouse has other features that are not described in the previous two sections, since they do not fall under any of MIC's affordances. One such feature is Clubhouse's calendar page, which displays upcoming rooms that are scheduled for each user. \newest{We were not able to uncover ways in which} the calendar feature enables anti-social behavior or promotes pro-social behavior, or aids in moderating the platform. As such, it is omitted \newest{in our representation} using MIC, allowing us to focus on just the features that are relevant.}

\subsubsection*{\newest{Accounting for Nuances at the Platform Level}}\newest{Though \citet{jiang19} uncovered the challenges that come with ``real-time, ephemeral'' voice communities, our case studies have shown that other platforms that also have \textbf{synchronous} and \textbf{ephemeral} audio might not face the same problems as those found in Discord. Namely, \citet{jiang19} found that moderators on Discord have a tough time gathering evidence of antisocial behavior since they are not present in voice channels at all times; this creates challenges since it limits the moderators' ability to exclude bad actors, and exclusion is one of Grimmelmann's four techniques for moderation. Guided by MIC, we found that Clubhouse rooms always have moderators present (\textbf{user roles, mechanisms}). Thus, potential moderation challenges on Clubhouse may not be due to the lack of mechanisms for exclusion, as it appears to be on Discord. Alternatively, it could be the case that there are different, novel consequences of synchronous, ephemeral voice settings that hinders moderators' ability to exclude bad actors. 
% For example, Clubhouse moderators might feel uncomfortable confronting bad actors since their user profiles are associated with their true identity.
}

\newest{We believe that MIC-derived insights, which can be explored in future studies regarding moderation on Clubhouse or synchronous, ephemeral audio platforms at large, are far more comprehensive than those that could be derived from Grimmelmann's view of moderation \cite{grimm}. In fact, \citet{jiang19} ultimately argue that ``designers and moderators should not ignore the technological infrastructure of the communit'' when trying to implement existing moderation strategies to new communities, and that they should ``carefully consider the limitations'' imposed by new infrastructure. Analysis via MIC begins with understanding the technological infrastructure of communities as it spells out the moderation-specific aspects of the platforms they are situated on. Having these insights could aid in future study design, and make the road to uncovering and extending findings, such as the ones from \citet{jiang19}, clearer.}

\subsubsection*{Understanding how Platform Changes effect Moderation} \newer{Another benefit of using MIC is that it let us pinpoint how specific changes on a platform could impact moderation (\Cref{sec:case-1}). Furthermore, we were able to use the inter-affordance relationships identified in MIC to get a more complete understanding of potential ways in which certain updates could effect Clubhouse's moderation ecosystem. For example, Clubhouse's new text-based messaging feature caused us to update the \textbf{modalities} affordance. However, since we used MIC to analyze Clubhouse, we observed that users used other platforms in tandem with Clubhouse to message one another (and therefore, a relationship between the \textbf{modalities} affordance and the \textbf{inter-platform} affordance). Thus, we could consider the possibility that a change to the \textbf{modalities} affordance would result in a change to \textbf{inter-platform} affordance. Using this inter-affordance relationship, we discussed potential impacts the above change might have had to moderation on Clubhouse. Without MIC, we might have overlooked this relationship and failed to investigate inter-platform relationships after modality changes.}

\newwriting{Additionally, changes on Clubhouse occurred over a period of six months, which is as long as a revision cycle in publication venues like CSCW. This means that moderation research and proposed moderation tools may become out-dated or obsolete more quickly. Using MIC as a common foundation with which to discuss moderation on social platforms would allow us to easily adapt and discuss how changes and updates to a platform may impact results of research and design.}

\subsubsection*{Adapting Moderation Mechanisms from Other Platforms}

\newwriting{\Cref{sec:case-2} demonstrates how MIC can be used to compare platforms in a systematic manner. MIC allows us to be mindful of how similar features across platforms can actually be impacted by different affordances. For instance, while Clubhouse and Spotify both have non-ephemeral content, Spotify's content is created asynchronously, while Clubhouse's content is created synchronously. Hence, while both platforms can moderate such content after-the-fact, Clubhouse has additional measures to ensure safety in live rooms. It is unclear to what additional moderation mechanisms Clubhouse has, if any, for its non-ephemeral content, apart from those listed in its Terms of Service. However, in comparing Spotify and Clubhouse, we could propose potential mechanisms for Clubhouse that are inspired by affordances in Spotify. Similarly, we used the comparisons between Discord and Clubhouse to propose moderation mechanisms for each platform that are inspired by each other.}

% \subsubsection*{Accessibility and Efficiency of MIC} Clubhouse is a novel platform that was difficult to approach and navigate. However, MIC allowed us to create a concise yet complete representation of its moderation ecosystem. It then allowed us to update that representation to Clubhouse that contained information that was relevent to moderation. We were also able to efficiently construct MIC diagrams for Discord and Spotify using observations and prior work, which allowed us to concisely describe the differences between these three platforms. The accessibility of MIC will make navigating ABSPs in general easier, and will provide a concise and standardized way to represent ABSPs. We believe that this will make moderation research for ABSPs easier to approach and unify.

% \subsubsection*{Connecting Relevant Research From Other Domains}  
% There is a decent amount of research done on moderation mechanisms of social platforms that fall outside the scope of traditional moderation research. For instance, Spotify uses music recommendations \cite{tang2017evaluating}, discovery \cite{aguiar2018platforms}, and curation mechanisms \cite{morris2015control, brovig2021remix} as organizational forms of moderation. Analysis via MIC will direct us to these non-moderation specific works, \newwriting{since we view moderation through the lens of platform affordances.}

\subsubsection*{Broader Effects of Affordance Changes} In \Cref{sec:case-2-mm}, we discussed potential moderation mechanisms that Discord, Spotify, and Clubhouse could adapt from one another. One proposal we made involved adapting Clubhouse's rule of keeping all recordings for a short period of time to address voice moderation challenges found on Discord \cite{jiang19}. We briefly discussed that users of Discord may not be open to this platform change, largely due to the fact that Discord seems to allow its users more privacy than Clubhouse does. This conjecture was made by observing that Discord users are allowed to be pseudonymous, while Clubhouse users have always been required to be identifiable. Observations like this are seemingly unimportant, and had we not used MIC, may have been overlooked. However, in some cases, \newest{design decisions that overlooked these subtle nuances directly preceded the downfall of platforms.}
%overlooking these subtle nuances have inadvertently allowed for detrimental platform changes.}

\newer{An example of this can be seen with YikYak, a social platform that allowed users to post location-specific anonymous text-posts~\cite{schlesinger17}. YikYak was a successful social platform that shut down in 2017 after platform changes were introduced. One such update was the removal of anonymity. As discussed in \Cref{sec:relwork}, existing research has explored the role anonymity played in voice-based interactions in online games \cite{wadley2015voice}. In particular, \citet{wadley2015voice} found that voice seemed to remove a degree of anonymity in game-play, which made some players feel uncomfortable, and in some cases, caused the players to abandon the game. \newest{We cannot retroactively prove that }
%There is no way to prove that 
MIC-based analysis would have prevented platform designers from making this platform change, \newest{nor can we assert that this specific change was the reason users abandoned the platform.} However, MIC would have highlighted anonymity as an integral affordance, and one that was similar to that of the online games and gaming platforms explored by \citet{wadley2015voice}. MIC-based analysis could have made these connections to a seemingly unrelated platform clear, and could have shed light on potential \newest{challenges that could result from altering the platform's anonymity affordance.} 
%from  (and later realized) pitfalls that could result from modifying the anonymity affordance. 
As such, MIC-based approach to moderation research and social platform design could be \newest{a valuable tool} in designing and maintaining successful social platforms.}

\subsection{Limitations and Future Work}

\subsubsection*{Limitations of MIC} \newwriting{MIC's purpose is for capturing the moderation ecosystems of social platforms to allow moderation researchers and platform stakeholders to better understand moderation. However, MIC does not capture every moderation-related property. In particular, the implicit norms that exist on a platform would not be represented by the affordances or relationships in MIC, since they are not tangible. Norms of online communities play a massive role in moderation on platforms, and is identified as one of four main moderation techniques by \citet{grimm}; there is also research that explores how norms play a role in moderating online communities, and how norms differ amongst various communities on the same platform \cite{chandrasekharan18norms,seering17}.}

\newwriting{Another closely related limitation of MIC is that it is not currently designed for analyzing the individual communities on each social platform. Studying individual online communities, such as specific subreddits, is beneficial for understanding moderation since each community has its own unique norms \cite{gilbert2020run}. There might be a way to extend MIC to capture nuances of community norms, which could be explored in future work.}

\subsubsection*{Extending MIC} MIC's base set of affordances and relationships are likely to become non-exhaustive as technology advances. Luckily, the graphical nature of MIC allows us to do so in an easy and straightforward way. We can add new affordances to our original set when new types of affordances that effect moderation are uncovered or developed. Similarly, we could further granularize existing affordances. For instance, we may eventually find it useful to distinguish between automated moderation mechanisms and manual ones. We can also extend our set of relationships by defining new types of relationships. There is no real restriction on how one could go about defining new relationships. We could even forego the condition that relationships occur between only two affordances, and describe multi-affordance relationships that are analogous to \emph{hyper-edges}.\footnote{In graph theory, a hyper-edge represents an ``edge'' that occurs between more than two vertices, usually represented as a subset of vertices.}

Another potentially useful, albeit more involved, extension of MIC, and in particular the MIC diagram would be to \newwriting{use the} inter-platform relationship affordance with a MIC diagram for other platforms or services. This would be useful if there is a nearly symbiotic relationship between two separate platforms or services, \newwriting{but we still wish to consider the affordances of each separately. For instance, Discord developed a new Clubhouse-like addition called Discord Stages\footnote{https://discord.com/stages}. It may be useful to consider Stages as a separate service from Discord's servers, since its use-case and set-up is different. We could analyze each of these %three 
services separately, and then build an extended MIC diagram to understand moderation on Discord in more detail.}

% \subsubsection*{Implications using MIC} \newwriting{MIC made it easy for us to generate these moderation interventions and reason about potential pitfalls they might cause. Without MIC, it would be challenging to not only see the areas in which these platforms and their designers could learn and adopt ideas from one another, but also identify other affordances that could be impacted if some such mechanism is implemented. Similar to in \Cref{sec:case-1}, the ease with which MIC can be updated to reflect new features and additions to platforms can make it easier for moderation researchers to conduct analyses of moderation of multiple platforms, and build upon previous, possibly out-dated, work. }

\section{Conclusion}

In this paper, we introduced the MIC framework as an \newest{affordance-aware augmentation of Grimmelmann's taxonomy \cite{grimm}, a popular lens for discussing moderation}. MIC provides a standardized way to represent moderation ecosystems of social platforms that \newest{highlights the moderation-related platform affordances involved, as well as the relationships that exist between them.} \newest{Over the course of two case studies, we used MIC to analyze a rapidly evolving platform (Clubhouse) and subsequently compare it to other relevant platforms (Discord and Spotify) to help generate possible new moderation interventions to address the challenges each may face.} We believe that the \newest{dynamic and comprehensive nature of the} MIC framework will help the moderation research community \newest{and moderation stakeholders effectively} keep up with the fast-paced nature of social platform development.

\bibliographystyle{ACM-Reference-Format}
\bibliography{references.bib}

% \appendix









\end{document}